\documentclass[twoside,draftcls,onecolumn]{IEEEtran}
\usepackage[utf8]{inputenc}
\usepackage{amsthm}
\usepackage{amsmath}
\usepackage{amssymb}
\usepackage{hyperref}

\makeatletter
\theoremstyle{plain}
\newtheorem{thm}{\protect\theoremname}

\theoremstyle{definition}
\newtheorem{defn}{\protect\definitionname}
\newtheorem{rem}{Remark}

\theoremstyle{plain}
\newtheorem{prop}[thm]{\protect\propositionname}

\usepackage{eurosym}
\usepackage{amsfonts}
\usepackage{graphicx}
\usepackage{epstopdf}
\usepackage{algpseudocode}
\usepackage{algorithmicx}
\usepackage{cases}
\usepackage{longtable}
\usepackage[usenames,dvipsnames]{color}
\usepackage{colortbl}
\usepackage{multirow}
\usepackage{verbatim}
\usepackage{placeins}
\usepackage{steinmetz}
\usepackage{cite}
\usepackage{enumitem}
\usepackage{placeins}
\usepackage[normalem]{ulem}

\makeatother

\providecommand{\definitionname}{Definition}
\providecommand{\propositionname}{Proposition}
\providecommand{\theoremname}{Theorem}

\begin{document}

\title{\textbf{The Marginalized $\delta$-GLMB Filter}}
\author{Claudio Fantacci, Ba-Tuong Vo, Francesco Papi and Ba-Ngu Vo\thanks{Ba-Tuong Vo, Francesco Papi and Ba-Ngu Vo are with the Department of Electrical and Computer Engineering, Curtin University, Bentley, WA, 6102, Australia (e-mail: \href{mailto:francesco.papi@curtin.edu.au}{francesco.papi@curtin.edu.au}, \href{mailto:ba-tuong@curtin.edu.au}{ba-tuong@curtin.edu.au}, \href{mailto:ba-ngu.vo@curtin.edu.au}{ba-ngu.vo@curtin.edu.au}).}\thanks{Claudio Fantacci is with the Dipartimento di Ingegneria dell’Informazione, Universit\`a di Firenze, Florence, 50139, Italy, on leave at the Curtin University, Bentley, WA, 6102, Australia, in the period January-July 2014 (e-mail: \href{mailto:claudio.fantacci@unifi.it}{claudio.fantacci@unifi.it}).}}

\maketitle
\begin{abstract}
The multi-target Bayes filter proposed by Mahler is a principled solution to recursive Bayesian tracking based on RFS or FISST. The $\delta$-GLMB filter is an exact closed form solution to the multi-target Bayes recursion which yields joint state and label or trajectory estimates in the presence of clutter, missed detections and association uncertainty. Due to presence of explicit data associations in the $\delta$-GLMB filter, the number of components in the posterior grows without bound in time. In this work we propose an efficient approximation to the $\delta$-GLMB filter which preserves both the PHD and cardinality distribution of the labeled posterior. This approximation also facilitates efficient multi-sensor tracking with detection-based measurements. Simulation results are presented to verify the proposed approach.
\end{abstract}


\section{Introduction}
\label{sec:intro}
Multi-target filtering/tracking involves the simultaneous estimation
of the number of targets along with their states, based on a sequence
of noisy measurements such as radar or sonar waveforms \cite{Mahler07}.
To reduce complexity and facilitate tractability, the sensor waveforms
are typically processed into a sequence of detections. The key challenges
in multi-target filtering/tracking thus include \textit{detection uncertainty},
\textit{clutter}, and \textit{data association uncertainty}. To date,
three major approaches to multi-target tracking/filtering have emerged
as the main solution paradigms. These are, Multiple Hypotheses Tracking
(MHT), \cite{Reid77,Kurien90,Blackman,Mallick12}, Joint Probabilistic
Data Association (JPDA) \cite{Blackman,Bar88}, and Random Finite
Set (RFS) \cite{Mahler07}.

The RFS or FInite Set STatistics (FISST) approach
pioneered by Mahler provides principled recursive Bayesian formulation
of the multi-target filtering/tracking problem. The essence of the
RFS approach is the modeling of the collection of target states and
measurements, referred to as the multi-target state and multi-target
measurement, as finite set valued random variables \cite{MahlerPHD2,Mahler07}.
The centerpiece of the RFS approach is the \textit{Bayes multi-target
filter} \cite{Mahler07}, which recursively propagates the filtering
density of the multi-target state forward in time. The PHD \cite{mahler2,vo-ma}, CPHD \cite{mahler2,vo-vo-cantoni} and
cardinality-balanced and labeled Multi-Bernoulli filters \cite{Vo2009,lmbf} are tractable
approximations to the Bayes multi-target filter which are synonymous
with the RFS framework. Their tractability however largely hinges
on the approximate form for the posterior which cannot accommodate
statistical dependencies between targets. 

The Bayes multi-target filter is also a (multi-target) tracker when
target identities or labels are incorporated into individual target
states. In \cite{VoVo11,VoConj13}, the notion of \textit{labeled RFSs}
is introduced to address target trajectories and their uniqueness.
The key results include conjugate priors that are closed under the
Chapman-Kolmogorov equation, and an analytic solution to the Bayes
multi-target tracking filter known as the $\delta$-Generalized Labeled
Multi-Bernoulli ($\delta$-GLMB) filter \cite{vovo2}. With detection
based measurements, the computational complexity in the $\delta$-GLMB
filter is mainly due to the presence of explicit data associations.
For certain applications such as tracking with multiple sensors, partially
observable measurements or decentralized estimation, the application
of a $\delta$-GLMB filter may not be possible due to limited computational
resources. Thus cheaper approximations to the $\delta$-GLMB filter
are of practical significance in multi-target tracking. 

In this paper we present a new approximation to the $\delta$-GLMB
filter. Our result is based on the approximation proposed in \cite{GLMBapprox}
where it was shown that the GLMB distribution can be used to construct
a principled approximation to an arbitrary labeled RFS density that
matches the PHD and the cardinality distribution. We refer to the
resultant filter as a Marginalized $\delta$-GLMB (M$\delta$-GLMB)
filter since it can be interpreted as a \textit{marginalization over
the data associations}. The proposed filter is consequently computationally
cheaper than the $\delta$-GLMB filter while still preserving key
summary statistics of the multi-target posterior. Importantly the
M$\delta$-GLMB filter facilitates tractable multi-sensor multi-target
tracking. Unlike PHD/CPHD and Multi-Bernoulli based filters, the proposed
approximation accommodates statistical dependence between targets.
We also present an alternative derivation of the LMB filter proposed
in \cite{lmbf} based on the newly proposed M$\delta$-GLMB filter.
Simulations results verify the proposed approximation.

\section{Background and problem formulation}
\label{sec:back}
This section briefly presents background material
on multi-object filtering and labeled RFS, which form the basis for
the formulation of our multi-target tracking problem.

\subsection{Multi-object Estimation}
Suppose that at time $k$, there are $N(k)$ object states $x_{k,1},\ldots,x_{k,N(k)}$,
each taking values in a state space ${\mathcal{X}}$. In the random
finite set (RFS) framework, the \textit{multi-object state} at time
$k$ is represented by the finite set $X_{k}=\{x_{k,1},\ldots,x_{k,N(k)}\}$,
and the multi-object state space is the space of all finite subsets
of $\mathcal{X}$, denoted as ${\mathcal{F}}{\mathbf{(}}{\mathcal{X}}{\mathbf{)}}$.
An RFS is simply a random variable that take values the space ${\mathcal{F}}{\mathbf{(}}{\mathcal{X}}{\mathbf{)}}$
that does not inherit the usual Euclidean notion of integration and
density. Mahler's Finite Set Statistics (FISST) provides powerful
yet practical mathematical tools for dealing with RFSs \cite{Mahler07}
based on a notion of integration/density that is consistent with point
process theory \cite{VSD05}.

Let $\pi_{k}(\cdot|Z_{k})$ denote the \textit{multi-target posterior density} at time $k$, and $\pi_{k+1|k}(\cdot)$ denote the \textit{multi-target prediction density} to time k + 1 (formally $\pi_{k}(\cdot)$ and $\pi_{k+1|k}(\cdot)$ should be written respectively as $\pi_{k}(\cdot|Z_{0}, \dots, Z_{k-1}, Z_{k})$, and $\pi_{k+1|k}(\cdot|Z_{0}, \dots, Z_{k})$, but for simplicity the dependence on past measurements is omitted).
Then, the \textit{multi-target Bayes recursion} propagates $\pi_{k}(\cdot)$ in time \cite{Mahler07,MahlerPHD2}, according to the following update and prediction
\begin{align}
	\pi_{k}(X_{k}|Z_{k}) & =\dfrac{g_{k}(Z_{k}|X_{k})\pi_{k|k-1}(X_{k})}{\displaystyle\int g_{k}(Z_{k}|X)\pi_{k|k-1}(X)\delta X} \, ,\label{eq:MTBayesUpdate}\\
	\pi_{k+1|k}(X_{k}) & =\int f_{k|k-1}(X_{k}|X)\pi_{k-1}(X)\delta X,\label{eq:MTBayesPred}
\end{align}
where $f_{k|k-1}$ is the \textit{multi-object transition density} to time $k+1$, $g_{k}$ is the \textit{multi-object likelihood function} at time $k$, and the integral is a \textit{set integral} defined for any function $f:{\mathcal{F}}{\mathbf{(}}{\mathcal{X}}{\mathbf{)}}\rightarrow{\mathbb{R}}$ by
\begin{equation}
	\int f(X)\delta X=\sum_{i=0}^{\infty}\frac{1}{i!}\int f(\{x_{1},...,x_{i}\})d(x_{1},...,x_{i}).
\end{equation}
An analytic solution to the multi-object Bayes filter for labeled states and track estimation from the multi-object filtering density was given in \cite{vovo1}.

\subsection{Labeled RFS}
To perform tracking in the RFS framework we use the label RFS model
that incorporates a unique label in the object's state vector to identify
its trajectory \cite{Mahler07}. In this model, the single-object
state space ${\mathcal{X}}$ is a Cartesian product ${\mathbb{X}}{\mathcal{\times}}{\mathbb{L}}$,
where ${\mathbb{X}}$ is the feature/kinematic space and ${\mathbb{L}}$
is the (discrete) label space. A finite subset set ${\mathbf{X}}$
of ${\mathbb{X}}{\mathcal{\times}}{\mathbb{L}}$ has distinct labels
if and only if ${\mathbf{X}}$ and its labels $\{\ell:(x,\ell)\in{\mathbf{X}}\}$
have the same cardinality. An RFS on ${\mathbb{X}}{\mathcal{\times}}{\mathbb{L}}$
with distinct labels is called a \textit{labeled RFS} \cite{vovo1}.

For the rest of the paper, we use the standard inner product notation
$\left\langle f,g\right\rangle \triangleq\int f(x)g(x)dx$, and multi-object
exponential notation $h^{X}\triangleq\prod_{_{x\in X}}h(x)$, where
$h$ is a real-valued function, with $h^{\emptyset}=1$ by convention.
We denote a generalization of the Kroneker delta and the inclusion
function that take arbitrary arguments such as sets, vectors, etc,
by
\begin{eqnarray*}
\delta_{Y}(X) & \triangleq & \left\{ \begin{array}{l}
1,{\text{ if }}X=Y\\
0,{\text{ otherwise}}
\end{array}\right.\\
1_{Y}(X) & \triangleq & \left\{ \begin{array}{l}
1,{\text{ if }}X\subseteq Y\\
0,{\text{ otherwise}}
\end{array}\right.
\end{eqnarray*}
We also write $1_{Y}(x)$ in place of $1_{Y}(\{x\})$ when $X=\{x\}$.
Single-object states are represented by lowercase letters, e.g. $x$,
${\mathbf{x}}$ while multi-object states are represented by uppercase
letters, e.g. $X$, ${\mathbf{X}}$, symbols for labeled states and
their distributions are bolded to distinguish them from unlabeled
ones, e.g. ${\mathbf{x}}$, ${\mathbf{X}}$, ${\mathbf{\pi}}$, etc,
spaces are represented by blackboard bold e.g. ${\mathbb{X}}$, ${\mathbb{Z}}$,
${\mathbb{L}}$, etc.

An important class of labeled RFS is the generalized labeled multi-Bernoulli
(GLMB) family \cite{vovo1}, which is the basis of an analytic solution
to the Bayes multi-object filter \cite{vovo2}. Under the standard
multi-object measurement model, the GLMB is a conjugate prior that
is also closed under the Chapman-Kolmogorov equation. If we start
with a GLMB initial prior, then the multi-object prediction and posterior
densities at any time are also GLMB densities. Let ${\mathcal{L}}:{\mathbb{X}}{\mathcal{\times}}{\mathbb{L}}\rightarrow{\mathbb{L}}$
be the projection ${\mathcal{L}}((x,\ell))=\ell$, and $\Delta({\mathbf{X}})\triangleq$$\delta_{|{\mathbf{X}}|}(|{\mathcal{L}}({\mathbf{X}})|)$
denote the \textit{distinct label indicator}. A GLMB is a labeled RFS
on ${\mathbb{X}}{\mathcal{\times}}{\mathbb{L}}$ distributed according
to 
\begin{equation}
{\mathbf{\boldsymbol{\pi}}}({\mathbf{X}})=\Delta({\mathbf{X}})\sum_{c\in{\mathbb{C}}}w^{(c)}({\mathcal{L}}({\mathbf{X}}))\left[p^{(c)}\right]^{{\mathbf{X}}}\label{eq:GLMB}
\end{equation}
where ${\mathbb{C}}$ is a discrete index set, $w^{(c)}(L)$ and $p^{(c)}$
satisfy: 
\begin{eqnarray}
\sum_{L\subseteq{\mathbb{L}}}\sum_{c\in{\mathbb{C}}}w^{(c)}(L) & = & 1,\\
\int p^{(c)}(x,\ell)dx & = & 1.
\end{eqnarray}
The GLMB density ($\ref{eq:GLMB}$) can be interpreted as a mixture
of multi-object exponentials. Each term in ($\ref{eq:GLMB}$) consists
of a weight $w^{(c)}({\mathcal{L}}({\mathbf{X}}))$ that depends only
on the labels of ${\mathbf{X}}$, and a multi-object exponential $\left[p^{(c)}\right]^{{\mathbf{X}}}$
that depends on the entire ${\mathbf{X}}$. The PHD (or intensity
function) of the unlabeled version of generalized labeled multi-Bernoulli
RFS is given by
\[
v(x)=\sum_{c\in{\mathbb{C}}}\sum_{\ell\in{\mathbb{L}}}p^{(c)}(x,\ell)\sum_{L\subseteq{\mathbb{L}}}1_{L}(\ell)w^{(c)}(L)
\]

The Labeled Multi-Bernoulli (LMB) family is a special case of the
GLMB family with one term: 
\begin{eqnarray*}
p^{(c)}(x,\ell) & = & p^{(\ell)}(x)\\
w^{(c)}(L) & = & \prod\limits _{\ell\in{\mathbb{M}}}\left(1-r^{(\ell)}\right)\prod\limits _{\ell\in L}\frac{1_{{\mathbb{M}}}(\ell)r^{(\ell)}}{1-r^{(\ell)}}.
\end{eqnarray*}
where $\{(r^{(\ell)},p^{(\ell)})\}_{\ell\in{\mathbb{M}}}$, ${\mathbb{M\subseteq L}}$,
is a given set of parameters with $r^{(\ell)}$ representing the existence
probability of track $\ell$, and $p^{(\ell)}$ the probability density
of the kinematic state of track $\ell$ given its existence \cite{vovo1}.
Note that the index space ${\mathbb{C}}$ has only one element, in
which case the $(c)$ superscript is not needed.
The LMB family is the basis of the LMB filter, an effective approximation of the Bayes multi-target tracking filter, which is highly parallelizable and capable of tracking large number of targets \cite{lmbf}.
The LMB filter, however, is an approximation of the Bayes multi-target tracking filter which only preserves the unlabeled PHD of the multi-target posterior \cite{lmbf}.
The information lost (e.g. the cardinality distribution of multi-target posterior and the approximate construction of the individual tracks) can lead to poor performance in cases of low observability and/or signal-to-noise ratio (SNR) which will be demonstrated in Section \ref{sec:results}.

\subsection{The $\delta$-GLMB Filter}
\label{ssec:dglmb}
An efficient approach to multi-target tracking
was presented in \cite{vovo1} using a special form of the GLMB distribution
in eq. (\ref{eq:GLMB}) called $\delta$-GLMB, i.e.
\begin{align}
\boldsymbol{\pi}(\mathbf{X}) & =\Delta(\mathbf{X})\sum_{\left(I,\xi\right)\in\mathcal{F}\left(\mathbb{L}\right)\times\Xi}w^{\left(I,\xi\right)}\delta_{I}\left(\mathcal{L}\left(\mathbf{X}\right)\right)\left[p^{\left(\xi\right)}\right]^{\mathbf{X}},\label{eq:dglmb}\\
 & =\Delta(\mathbf{X})\sum_{I\in\mathcal{F}\left(\mathbb{L}\right)}\delta_{I}\left(\mathcal{L}\left(\mathbf{X}\right)\right)\sum_{\xi\in\Xi}w^{\left(I,\xi\right)}\left[p^{\left(\xi\right)}\right]^{\mathbf{X}}.\label{eq:dglmb2}
\end{align}
The $\delta$-GLMB density naturally arises in multi-target tracking
problems when using the standard detection based measurement model.
In the following we briefly recall the prediction and update steps
for the $\delta$-GLMB filter, additional details can be found in
\cite{vovo2}. To ensure distinct labels
we assign each target an ordered pair of integers $\ell=(k,i)$, where
$k$ is the time of birth and $i$ is a unique index to distinguish
targets born at the same time. The label space for targets born at
time $k+1$ is denoted as ${\mathbb{L}}_{k+1}$, and a target born
at time $k+1$, has state ${\mathbf{x}}\in{\mathbb{X}}{\mathcal{\times}}{\mathbb{L}}_{k+1}$.
The label space for targets up to time $k+1$ (i.e. including those born prior
to $k+1$), denoted as ${\mathbb{L}}_{0:k+1}$, is constructed recursively
by ${\mathbb{L}}_{0:k+1}={\mathbb{L}}_{0:k}\cup{\mathbb{L}}_{k+1}$
(note that ${\mathbb{L}}_{0:k}$ and ${\mathbb{L}}_{k+1}$ are disjoint).
A multi-object state ${\mathbf{X}}$ at time $k+1$, is a finite subset
of ${\mathcal{X=}}$ ${\mathbb{X}}{\mathcal{\times}}{\mathbb{L}}_{0:k+1}$.

Suppose that at time $k$, there are $N_{k}$ objects with states $\mathbf{x}_{k,1}, \dots, \mathbf{x}_{k, N_{k}}$, each taking values in the (labeled) state space $\mathbb{X} \times \mathbb{L}_{0:k}$, and $M_{k}$ measurements $z_{k,1}, \dots, z_{k, M_{k}}$ each taking values in an observation space $\mathbb{Z}$.
The \textit{multi-object state} and \textit{multi-object observation}, at time $k$, \cite{Mahler07,MahlerPHD2} are, respectively, the finite sets 
\begin{align}
	\mathbf{X}_{k} & = \left\{ \mathbf{x}_{k,1}, \dots, \mathbf{x}_{k,N_{k}} \right\} \, ,\\
	Y_{k} & = \left\{ y_{k,1}, \dots,y_{k,M_{k}} \right\} \, .
\end{align}
The $\delta$-GLMB filter recursively propagates a $\delta$-GLMB posterior density forward in time according to the following Bayesian update and prediction
\begin{align}
	\boldsymbol{\pi}_{k}(\mathbf{X}_{k}|Z_{k}) & =\dfrac{g_{k}(Z_{k}|\mathbf{X}_{k})\boldsymbol{\pi}_{k|k-1}(\mathbf{X}_{k})}{\displaystyle\int g_{k}(Z_{k}|\mathbf{X})\boldsymbol{\pi}_{k|k-1}(\mathbf{X})\delta \mathbf{X}} \, ,\label{eq:LMTBayesUpdate}\\
	\boldsymbol{\pi}_{k+1|k}(\mathbf{X}_{k}) & =\int f_{k|k-1}(\mathbf{X}_{k}|\mathbf{X}) \boldsymbol{\pi}_{k-1}(\mathbf{X})\delta \mathbf{X} \, ,\label{eq:LMTBayesPred}
\end{align}
which is the labeled counterpart of the Bayesian recursion (\ref{eq:MTBayesUpdate})-(\ref{eq:MTBayesPred}).

\subsubsection{$\delta$-GLMB Prediction}
Given the current multi-object state ${\mathbf{X}}^{\prime}$, each state
$(x^{\prime},\ell^{\prime})$ $\in{\mathbf{X}}^{\prime}$ either continues
to exist at the next time step with probability $P_{S}(x^{\prime},\ell^{\prime})$
and evolves to a new state $(x,\ell)$ with probability density $f_{k+1|k}(x|x^{\prime},\ell^{\prime})\delta_{\ell}(\ell^{\prime})$,
or dies with probability $1 - P_{S}(x^{\prime},\ell^{\prime})$.
Note that the label of the objects is preserved in the transition, only the kinematic part of state changes.
Assuming that $\mathbf{X}$ has distinct labels and that conditional on $\mathbf{X}$, the transition of the kinematic states are mutually independent, then the set $\mathbf{W}$ of surviving objects at the next time is a labeled multi-Bernoulli RFS \cite{vovo1}
\begin{equation}
	\mathbf{f}_{S}\!\left( \mathbf{W} | \mathbf{X} \right) = \Delta\!\left( \mathbf{W} \right) \Delta\!\left( \mathbf{X} \right) \, 1_{\mathcal{L}\left( \mathbf{X} \right)}\left( \mathcal{L}\!\left( \mathbf{W} \right) \right) \left[ \Phi\left( \mathbf{W}; \cdot \right) \right]^{\mathbf{X}} \, ,\label{eq:survivorpdf}
\end{equation}
where
\begin{equation}
	\Phi\!\left( \mathbf{W}; x^{\prime}, \ell^{\prime} \right) = \sum_{\left( x, \ell \right) \in \mathbf{W}} \delta_{\ell^{\prime}}\left( \ell \right) P_{S}\left( x^{\prime}, \ell^{\prime} \right) f\left( x | x^{\prime}, \ell^{\prime} \right) + \left[ 1 - 1_{\mathcal{L}\left( \mathbf{X} \right)}\left( \ell^{\prime} \right) \right] \left( 1 - P_{S}\left( x^{\prime}, \ell^{\prime} \right) \right) \, .
\end{equation}
The $\Delta\!\left( \mathbf{X} \right)$ in (\ref{eq:survivorpdf}) ensures that only $\mathbf{X}$ with distinct labels are considered.
The set of new objects born at the next time step is distributed according to 
\begin{equation}
{\mathbf{f}}_{B}({\mathbf{Y}})=\Delta({\mathbf{Y}})w_{B}({\mathcal{L}}({\mathbf{Y}}))\left[p_{B}\right]^{{\mathbf{Y}}}\label{eq:Birth_transition}
\end{equation}
The birth density ${\mathbf{f}}_{B}$ is defined on ${\mathbb{X}}\times{\mathbb{L}}_{k+1}$
and ${\mathbf{f}}_{B}({\mathbf{Y}})=0$ if ${\mathbf{Y}}$ contains
any element ${\mathbf{y}}$ with ${\mathcal{L}}(\mathbf{y})\notin{\mathbb{L}}_{k+1}$.
The birth model (\ref{eq:Birth_transition}) covers both labeled Poisson
and labeled multi-Bernoulli.
The multi-object state at the next time $\mathbf{X}$ is the superposition of surviving objects and new born objects, i.e. $\mathbf{X} = \mathbf{W} \cup \mathbf{Y}$.
Since the label spaces $\mathbb{L}$ and $\mathbb{B}$ are disjoint, the labeled birth objects and surviving objects are independent.
Thus the multi-target transition density turns out to be the product of the transition density (\ref{eq:survivorpdf}) and the density of new objects (\ref{eq:Birth_transition})
\begin{equation}
	\mathbf{f}\!\left( \mathbf{X} | \mathbf{X}^{\prime} \right) = \mathbf{f}_{S}\!\left( \mathbf{X} \cap \left( \mathbb{X} \times \mathbb{L} \right) | \mathbf{X}^{\prime} \right) \, \mathbf{f}_{B}\left( \mathbf{X} - \left( \mathbb{X} \times \mathbb{L} \right) \right) \, .
\end{equation}
Additional details can be found in \cite{vovo1}.

If the current multi-object prior density is a $\delta$-GLMB of the form (\ref{eq:dglmb}), then the multi-object
prediction density is a $\delta$-GLMB given by
\begin{equation}
	{\mathbf{\boldsymbol{\pi}}}_{k+1|k}({\mathbf{X}})=\Delta({\mathbf{X}})\sum_{\left(I,\xi\right)\in{\mathcal{{\mathcal{F}}}}({\mathbb{{\mathbb{L}}}}_{0:k+1})\times\Xi}w_{k+1|k}^{\left(I,\xi\right)}\delta_{I}\left(\mathcal{L}\left(\mathbf{X}\right)\right)\left[p_{k+1|k}^{\left(\xi\right)}\right]^{\mathbf{X}}\label{eq:dglmbpredictedpdf}
\end{equation}
where
\begin{eqnarray}
w_{k+1|k}^{\left(I,\xi\right)} & = & w_{B}(I-{\mathbb{L}}_{0:k})w_{S}^{(\xi)}(I\cap{\mathbb{L}}_{0:k}) \, ,\\
p_{k+1|k}^{(\xi)}(x,\ell) & = & 1_{{\mathbb{L}}_{0:k}}(\ell)p_{S}^{(\xi)}(x^{\prime},\ell)+1_{{\mathbb{L}}_{k+1}}(\ell)p_{B}(x,\ell) \, ,\\
p_{S}^{(\xi)}(x^{\prime},\ell) & = & \frac{\left\langle P_{S}(\cdot,\ell)f_{k+1|k}(x|\cdot,\ell),p_{k}^{(\xi)}(\cdot,\ell)\right\rangle }{\eta_{S}^{(\xi)}(\ell)} \, ,\\
\eta_{S}^{(\xi)}(\ell) & = & \left\langle P_{S}(\cdot,\ell),p_{k}^{(\xi)}(\cdot,\ell)\right\rangle \, ,\\
w_{S}^{(\xi)}(L) & = & [\eta_{S}^{(\xi)}]^{L}\sum_{J\subseteq{\mathbb{L}}_{0:k}}1_{J}(L)[1-\eta_{S}^{(\xi)}]^{J-L}w_{k}^{(I,\xi)} \, .
\end{eqnarray}

\subsubsection{$\delta$-GLMB Update}
The standard multi-object observation model is described as follows.
For a given multi-object state $\mathbf{X}$, each state $\mathbf{x}\in\mathbf{X}$
is either detected with probability $P_{D}\left(\mathbf{x}\right)$
and generates a point $z$ with likelihood $g(z|\mathbf{x})$, or
missed with probability $1 - P_{D}\left(\mathbf{x}\right)$,
i.e. $\mathbf{x}$ generates a Bernoulli RFS with parameter $(P_{D}(\mathbf{x}),g(\cdot|\mathbf{x}))$.
Assuming that conditional on $\mathbf{X}$ these Bernoulli RFSs are
independent, then the set $W\subset\mathbb{Z}$ of detected points
(non-clutter measurements) is a multi-Bernoulli RFS with parameter
set $\{(P_{D}(\mathbf{x}),g(\cdot|\mathbf{x}))$: $\mathbf{x}\in\mathbf{X}\}$.
The set $Y\subset\mathbb{Z}$ of false observations (or clutter),
assumed independent of the detected points, is modeled by a Poisson
RFS with intensity function $\kappa(\cdot)$. The multi-object observation
$Z$ is the superposition of the detected points and false observations,
i.e. $Z=W\cup Y$, and the multi-target likelihood can be derived
as shown in \cite{vovo1}.
Assuming that, conditional on $\mathbf{X}$, detections are independent, and that clutter is independent of the detections, the multi-object likelihood is given by 
\begin{equation}
	g_{k}(Y|\mathbf{X})=e^{-\left\langle \kappa ,1\right\rangle}\kappa^{Y}\sum_{\theta \in \Theta (\mathcal{L}(\mathbf{X}))}\left[\psi_{Y}(\cdot ;\theta )\right] ^{\mathbf{X}} \, ,\label{eq:RFSmeaslikelihood0}
\end{equation}
where $\Theta (I)$ is the set of mappings $\theta:I \rightarrow \{0,1,...,M\},$ such that $\theta(i) = \theta( i^{\prime} ) > 0$ implies $i = i^{\prime}$, and
\begin{equation}
	\psi_{Y}(x, \ell; \theta ) =	\left\{ 
							\begin{array}{ll}
								\dfrac{P_{D}(x,\ell) \, g_{k}(y_{\theta(\ell)}|x,\ell)}{\kappa (y_{\theta(\ell)})} \, , & \text{if }\theta (\ell) > 0 \\ 
								1-P_{D}(x,\ell) \, , & \text{if }\theta(\ell) = 0
							\end{array} \right. \, .\label{eq:molikelihood}
\end{equation}
Note that an association map $\theta$ specifies which tracks generated which measurements, i.e. track $\ell$ generates measurement $y_{\theta(\ell)}\in Y$, with undetected tracks assigned to $0$.
The condition ``$\theta(i)=\theta (i^{\prime}) > 0$ implies $i = i^{\prime}$'', means that a track can generate at most one measurement, and a measurement can be assigned to at most one track, at one time instant.

If the current multi-object prediction density
is a $\delta$-GLMB of the form (\ref{eq:dglmb}), then the multi-object
posterior density is a $\delta$-GLMB given by
\begin{equation}
\boldsymbol{\pi}_{k}\left(\mathbf{X}|Z\right)=\Delta(\mathbf{X})\sum_{\left(I,\xi\right)\in\mathcal{F}\left(\mathbb{L}\right)\times\Xi}\sum_{\theta\in\Theta(I)}w_{k}^{\left(I,\xi,\theta\right)}(Z)\delta_{I}\left(\mathcal{L}\left(\mathbf{X}\right)\right)\left[p_{k}^{\left(\xi,\theta\right)}\left(\cdot|Z\right)\right]^{\mathbf{X}}\label{eq:dglmbupdatedpdf}
\end{equation}
where $\Theta(I)$ denotes the subset of the current maps with domain
$I$, and
\begin{align}
w_{k}^{(I,\xi,\theta)} & \propto w_{k+1|k}^{\left(I,\xi\right)}\left[\eta_{Z}^{(\xi,\theta)}(\ell)\right]^{I} \, ,\label{eq:updateweight}\\
\eta_{Z}^{(\xi,\theta)}(\ell) & =\left\langle p_{k+1|k}^{(\xi)}(\cdot,\ell),\psi_{Z}(\cdot,\ell;\theta)\right\rangle \, ,\nonumber \\
p_{k}^{\left(\xi,\theta\right)}\left(\cdot|Z\right) & =\frac{p_{k+1|k}^{(\xi)}(x,\ell)\psi_{Z}(x,\ell;\theta)}{\eta_{Z}^{(\xi,\theta)}(\ell)} \, .\nonumber
\end{align}
Notice that the new association maps $\theta$ can be added (stacked) to their respective association histories $\xi$ in order to have again the more compact form (\ref{eq:dglmb}) for the updated $\delta$-GLMB (\ref{eq:dglmbupdatedpdf}).

\section{The Marginalized $\delta$-GLMB Filter}
\label{sec:marginal}
In this section we present a new solution for recursive multi-target tracking based on the GLMB approximation technique presented in \cite{GLMBapprox}.
The resultant filter is called the Marginalized $\delta$-GLMB (M$\delta$-GLMB) filter since the result can be interpreted as performing a marginalization with respect to the association histories.
We present two important justifications for the new algorithm, namely, a computationally efficient approximation of the Bayes optimal $\delta$-GLMB filter which directly facilitates multi-sensor updates, and a theoretical result showing that the proposed approximation matches exactly the (labeled) PHD and the cardinality distribution of the filtering density.
Furthermore we show a connection with the LMB filter by presenting an alternative derivation of the LMB filter based on extracting individual tracks from the M$\delta$-GLMB filter.

\subsection{Marginalized $\delta$-GLMB Approximation}
One of the main factors contributing to the computational complexity of the $\delta$-GLMB filter \cite{vovo2} is the exponential growth of the number of hypotheses in the update of the prior (\ref{eq:updateweight}) which gives rise to the an explicit sum over an association history variable. Moreover, in multi-sensor scenarios the number of association histories is further increased due to successive update steps (as detailed in subsection \ref{sec:dmglmb-msensor}).
The idea behind the proposed M$\delta$-GLMB filter is to construct a principled GLMB approximation $\hat{\boldsymbol{\pi}}\left(\mathbf{\cdot}\right)$ to the posterior density $\boldsymbol{\pi}\left(\cdot\right)$ which results in a marginalization over the association histories thereby drastically reducing the number of components required to represent the posterior or filtering density.
\begin{defn}
A Marginalized $\delta$-GLMB density $\hat{\boldsymbol{\pi}}$ corresponding
to the $\delta$-GLMB density $\boldsymbol{\pi}$ in (\ref{eq:dglmb})
is a probability density of the form
\begin{equation}
\hat{\boldsymbol{\pi}}(\mathbf{X})=\Delta(\mathbf{X})\sum_{I\in\mathcal{F}(\mathbb{L})}\delta_{I}(\mathcal{L}(\mathbf{X}))w^{(I)}\left[p^{(I)}\right]^{\mathbf{X}}\label{eq:mdglmb}
\end{equation}
where
\begin{align}
w^{(I)} & =\sum_{\xi\in\Xi}w^{(I,\xi)} \, ,\label{eq:mdglmb_w}\\
p^{(I)}(x,\ell) & =1_{I}(\ell)\frac{1}{w^{(I)}}\sum_{\xi\in\Xi}w^{(I,\xi)}p^{(\xi)}(x,\ell) \, .\label{eq:mdglmb_p}
\end{align}
\end{defn}
\begin{prop}
\label{thm:mdglmb} The Marginalized $\delta$-GLMB density $\hat{\boldsymbol{\pi}}$
in (\ref{eq:mdglmb})-(\ref{eq:mdglmb_p}) preserves both PHD
and cardinality distribution of the original $\delta$-GLMB density
$\boldsymbol{\pi}$ in (\ref{eq:dglmb}).\end{prop}
\begin{IEEEproof}
We apply the result in Proposition 2 of \cite{GLMBapprox} which can
be used to calculate the parameters of the marginalized $\delta$-GLMB
density. Notice that the result in \cite{GLMBapprox} applies to any
labeled RFS density and our first step is to rewrite the $\delta$-GLMB
density (\ref{eq:dglmb}) in the general form for a labeled RFS density
specified in \cite{GLMBapprox}, i.e. $\boldsymbol{\pi}(\mathbf{X})=w(\mathcal{L}(\mathbf{X}))p(\mathbf{X})$
where 
\begin{align*}
w(\left\{ \ell_{1},\ldots,\ell_{n}\right\} ) & \triangleq\int_{\mathbb{X}^{n}}\boldsymbol{\pi}(\left\{ (x_{1},\ell_{1}),\ldots,(x_{n},\ell_{n})\right\} )d(x_{1},\ldots,x_{n})\\
 & =\sum_{I\in\mathcal{F}(\mathbb{L})}\delta_{I}(\left\{ \ell_{1},\ldots,\ell_{n}\right\} )\sum_{\xi\in\Xi}w^{(I,\xi)}\int_{\mathbb{X}^{n}}p^{(\xi)}(x_{1},\ell_{1})\cdots p^{(\xi)}(x_{n},\ell_{n})dx_{1}\cdots dx_{n}\\
 & =\sum_{\xi\in\Xi}w^{(\left\{ \ell_{1},\ldots,\ell_{n}\right\} ,\xi)}\sum_{I\in\mathcal{F}(\mathbb{L})}\delta_{I}(\left\{ \ell_{1},\ldots,\ell_{n}\right\} )\\
 & =\sum_{\xi\in\Xi}w^{(\left\{ \ell_{1},\ldots,\ell_{n}\right\} ,\xi)}
\end{align*}
and
\begin{align}
	p(\left\{ (x_{1},\ell_{1}),\ldots,(x_{n},\ell_{n})\right\} ) & \triangleq\frac{\boldsymbol{\pi}(\left\{ (x_{1},\ell_{1}),\ldots,(x_{n},\ell_{n})\right\} )}{w(\left\{ \ell_{1},\ldots,\ell_{n}\right\} )}\nonumber \\
	& =\Delta(\left\{ (x_{1},\ell_{1}),\ldots,(x_{n},\ell_{n})\right\} )\dfrac{\displaystyle\sum_{I\in\mathcal{F}\left(\mathbb{L}\right)}\delta_{I}\left(\left\{ \ell_{1},\ldots,\ell_{n}\right\} \right)\sum_{\xi\in\Xi}w^{\left(I,\xi\right)}\,\left[p^{\left(\xi\right)}\right]^{\left\{ (x_{1},\ell_{1}),\ldots,(x_{n},\ell_{n})\right\} }}{w(\left\{ \ell_{1},\ldots,\ell_{n}\right\} )}\label{eq:p}
\end{align}
Applying Proposition 2 of \cite{GLMBapprox}, the parameters $w^{(I)}$
and $p^{(I)}$ for the M$\delta$-GLMB approximation that match the
cardinality and PHD are
\begin{align*}
w^{(I)}(L) & =\delta_{I}(L)w(I)=\delta_{I}(L)\sum_{\xi\in\Xi}w^{(I,\xi)}
\end{align*}
and
\begin{align}
p^{(I)}(x,\ell) & =1_{I}(\ell)p_{I-\left\{ \ell\right\} }(x,\ell)\nonumber\\
 & =1_{I}(\ell)\int p(\left\{ (x,\ell),(x_{1},\ell_{1}),\ldots,(x_{j},\ell_{j})\right\} )d(x_{1},\ldots,x_{j})\label{eq:pmarignal}
\end{align}
where we enumerate $I-\{\ell\}=\left\{ \ell_{1},\ldots,\ell_{j}\right\} $.
Substituting the expression (\ref{eq:p}) in (\ref{eq:pmarignal}) we have
\begin{align*}
	p^{(I)}(x,\ell) & =1_{I}(\ell)\Delta(\left\{ (x,\ell),(x_{1},\ell_{1}),\ldots,(x_{j},\ell_{j})\right\} ) \dfrac{1}{w(\left\{ \ell,\ell_{1},\ldots,\ell_{j}\right\} )}\cdot\\
	& \qquad\qquad\qquad\qquad \cdot \sum_{J\in\mathcal{F}\left(\mathbb{L}\right)}\delta_{J}\left(\left\{ \ell,\ell_{1},\ldots,\ell_{j}\right\} \right) \sum_{\xi\in\Xi}w^{\left(J,\xi\right)}\,\int\left[p^{\left(\xi\right)}\right]^{\left\{ (x,\ell),(x_{1},\ell_{1}),\ldots,(x_{j},\ell_{j})\right\} }d(x_{1},\ldots,x_{j})\\
	& =1_{I}(\ell)\Delta(\left\{ (x,\ell),(x_{1},\ell_{1}),\ldots,(x_{j},\ell_{j})\right\}) \frac{1}{w(\left\{ \ell,\ell_{1},\ldots,\ell_{j}\right\} )} \sum_{J\in\mathcal{F}\left(\mathbb{L}\right)}\delta_{J}\left(\left\{ \ell,\ell_{1},\ldots,\ell_{j}\right\} \right)\sum_{\xi\in\Xi}w^{\left(J,\xi\right)}p^{(\xi)}(x,\ell)
\end{align*}
and noting that $I=\left\{ \ell,\ell_{1},\ldots,\ell_{j}\right\} $
it follows that only one term in the sum over $J$ is non-zero thus
giving
\[
p^{(I)}(x,\ell)=1_{I}(\ell)\Delta(\left\{ (x,\ell),(x_{1},\ell_{1}),\ldots,(x_{j},\ell_{j})\right\} )\frac{1}{\displaystyle\sum_{\xi\in\Xi}w^{(I,\xi)}}\sum_{\xi\in\Xi}w^{\left(I,\xi\right)}p^{(\xi)}(x,\ell)
\]
Consequently, the M$\delta$-GLMB approximation is given by
\begin{align*}
\hat{\boldsymbol{\pi}}(\mathbf{X}) & =\sum_{I\in\mathcal{F}(\mathbb{L})}w^{(I)}(\mathcal{L}(\mathbf{X}))\left[p^{(I)}\right]^{\mathbf{X}}\\
 & =\Delta(\mathbf{X})\sum_{I\in\mathcal{F}(\mathbb{L})}\delta_{I}(\mathcal{L}(\mathbf{X}))\sum_{\xi\in\Xi}w^{(I,\xi)}\left[1_{I}(\cdot)\frac{1}{\displaystyle\sum_{\xi\in\Xi}w^{(I,\xi)}}\sum_{\xi\in\Xi}w^{\left(I,\xi\right)}p^{(\xi)}(\cdot,\cdot)\right]^{\mathbf{X}}\\
 & =\Delta(\mathbf{X})\sum_{I\in\mathcal{F}(\mathbb{L})}\delta_{I}(\mathcal{L}(\mathbf{X}))w^{(I)}\left[p^{(I)}\right]^{\mathbf{X}}
\end{align*}
where
\begin{align*}
w^{(I)} & =\sum_{\xi\in\Xi}w^{(I,\xi)}\\
p^{(I)}(x,\ell) & =1_{I}(\ell)\frac{1}{\displaystyle\sum_{\xi\in\Xi}w^{(I,\xi)}}\sum_{\xi\in\Xi}w^{\left(I,\xi\right)}p^{(\xi)}(x,\ell)\\
 & =1_{I}(\ell)\frac{1}{w^{(I)}}\sum_{\xi\in\Xi}w^{\left(I,\xi\right)}p^{(\xi)}(x,\ell)
\end{align*}
\end{IEEEproof}


\subsection{M$\delta$-GLMB Recursion}
The M$\delta$-GLMB density can be exploited to construct an efficient recursive multi-object tracking filter by calculating the M$\delta$-GLMB approximation step after the $\delta$-GLMB update, and predicting forward in time using the $\delta$-GLMB prediction.

\subsubsection{M$\delta$-GLMB Prediction}
Given an updated density of the form (\ref{eq:mdglmb}) the M$\delta$-GLMB prediction step turns out to be
\begin{equation}
	{\mathbf{\boldsymbol{\pi}}}_{k+1|k}({\mathbf{X}}) = \Delta({\mathbf{X}}) \sum_{I \in \mathcal{F}(\mathbb{L}_{0:k+1})} w_{k+1|k}^{\left(I\right)}\delta_{I}\left(\mathcal{L}\left(\mathbf{X}\right)\right)\left[p_{k+1|k}^{\left(I\right)}\right]^{\mathbf{X}}\label{eq:mdglmbpredictedpdf}
\end{equation}
where
\begin{eqnarray}
w_{k+1|k}^{\left(I\right)} & = & w_{B}(I-{\mathbb{L}}_{0:k})w_{S}^{(I)}(I\cap{\mathbb{L}}_{0:k}) \, ,\\
p_{k+1|k}^{(I)}(x,\ell) & = & 1_{{\mathbb{L}}_{0:k}}(\ell)p_{S}^{(I)}(x^{\prime},\ell)+1_{{\mathbb{L}}_{k+1}}(\ell)p_{B}(x,\ell) \, ,\\
p_{S}^{(I)}(x^{\prime},\ell) & = & \frac{\left\langle P_{S}(\cdot,\ell)f_{k+1|k}(x|\cdot,\ell),p_{k}^{(I)}(\cdot,\ell)\right\rangle }{\eta_{S}^{(I)}(\ell)} \, ,\\
\eta_{S}^{(I)}(\ell) & = & \left\langle P_{S}(\cdot,\ell),p_{k}^{(I)}(\cdot,\ell)\right\rangle \, ,\\
w_{S}^{(I)}(L) & = & [\eta_{S}^{(I)}]^{L}\sum_{J\subseteq{\mathbb{L}}_{0:k}}1_{J}(L)[1-\eta_{S}^{(I)}]^{J-L}w_{k}^{(I)} \, ,
\end{eqnarray}
which is exactly the $\delta$-GLMB prediction step (\ref{eq:dglmbpredictedpdf}) with no association histories from previous time step, i.e. $\Xi = \varnothing$, and with the convention of having the superscript $(I)$ instead of $(\xi)$ due to the marginalization (\ref{eq:mdglmb_w})-(\ref{eq:mdglmb_p}).

\begin{rem}\label{rem:maxhppred}
	The number of components $\left( w_{k+1|k}^{\left(I\right)}, p_{k+1|k}^{\left(I\right)} \right)$ computed after the M$\delta$-GLMB prediction step (\ref{eq:mdglmbpredictedpdf}) is $\left|\mathcal{F}(\mathbb{L}_{0:k+1})\right|$.
	On the other hand, the number of components $\left( w_{k+1|k}^{\left(I, \xi\right)}, p_{k+1|k}^{\left(\xi\right)} \right)$ after the $\delta$-GLMB prediction (\ref{eq:dglmbpredictedpdf}) is $\left|\mathcal{F}(\mathbb{L}_{0:k+1}) \times \Xi \right|$ for $w_{k+1|k}^{\left(I, \xi\right)}$ and $\left| \Xi \right|$ for $p_{k+1|k}^{\left(\xi\right)}$.
	Notice that the number of weights $w_{k+1|k}^{\left(I\right)}$ of the M$\delta$-GLMB is substantially lower than the $w_{k+1|k}^{\left(I, \xi\right)}$ of $\delta$-GLMB.
	As for the number of location PDFs $p_{k+1|k}^{\left(\cdot\right)}$, it is worth noticing that the growth rate of the association histories $\xi \in \Xi$ is super-exponential with time \cite{vovo1,vovo2}, while the growth rate of the cardinality of $\mathcal{F}(\mathbb{L}_{0:k+1})$ is by far more restrained.
\end{rem}

The use of the M$\delta$-GLMB approximation further reduces the number of hypotheses in the posterior density while preserving the PHD and cardinality distribution \cite{GLMBapprox}.
Moreover, the M$\delta$-GLMB is in a form that it is suitable for efficient and tractable information fusion (i.e. multi-sensor processing) which will be shown in the next subsection.

\subsubsection{Multi-Sensor M$\delta$-GLMB Update}\label{sec:dmglmb-msensor}
Consider now a multi-sensor setting in which the sensors (indexed with $s$) convey all the measurement sets $Z^{s}$ to a central fusion node.
Assuming that such a measurement sets taken by the sensors are conditionally independent on the states, the multi-object Bayesian filtering update (\ref{eq:LMTBayesUpdate}) can be naturally extended as follows:
\begin{equation}
	\boldsymbol{\pi}(\mathbf{X}_{k}) = \dfrac{\displaystyle\prod_{s} g^{s}_{k}(Z^{s}_{k}|\mathbf{X}_{k}) \boldsymbol{\pi}_{k|k-1}(\mathbf{X}_{k})}{\displaystyle\int \prod_{\varsigma} g^{\varsigma}_{k}(Z^{\varsigma}_{k}|\mathbf{X}) \boldsymbol{\pi}_{k|k-1}(\mathbf{X})\delta \mathbf{X}} \, .\label{eq:msMTBayesUpdate}
\end{equation}
where $g^{s}_{k}$ is the multi-object likelihood of sensor $s$.
Thus, at each time instant $k$, the M$\delta$-GLMB update step (\ref{eq:mdglmbupdatedpdf}) (and equivalently for the $\delta$-GLMB update step (\ref{eq:dglmbupdatedpdf})) is sequentially repeated exploiting the measurement sets $Z^{s}_{k}$ provided by the sensors.

Let us now focus on the single update step to be carried out for each sensor $s$.
If the current multi-object prior density is a M$\delta$-GLMB of the form (\ref{eq:mdglmb}), then the multi-object posterior density is a $\delta$-GLMB given by
\begin{equation}
	\boldsymbol{\pi}_{k}\left(\mathbf{X}|Z\right) = \Delta(\mathbf{X})
	\sum_{I \in \mathcal{F}\left(\mathbb{L}\right)} \,
	\sum_{\theta \in \Theta(I)} w_{k}^{\left(I,\theta\right)}(Z) \delta_{I}\left(\mathcal{L}\left(\mathbf{X}\right)\right)\left[ p_{k}^{\left(I,\theta\right)}\left(\cdot|Z\right)\right]^{\mathbf{X}}\label{eq:mdglmbupdatedpdf}
\end{equation}
where $\Theta(I)$ (see (\ref{eq:RFSmeaslikelihood0})) denotes the subset of the current maps with domain $I$ , and
\begin{align}
	w_{k}^{(I,\theta)} & \propto w_{k+1|k}^{\left( I \right)}\left[\eta_{Z}^{( I, \theta )}(\ell)\right]^{I} \, ,\label{eq:mdglmbupdateweight}\\
	\eta_{Z}^{(I,\theta)}(\ell) & =\left\langle p_{k+1|k}^{(I)}(\cdot,\ell),\psi_{Z}(\cdot,\ell;\theta)\right\rangle \, , \\
	p_{k}^{\left( I, \theta \right)}\left( \cdot | Z \right) & = \frac{p_{k+1|k}^{(I)}(x,\ell) \, \psi_{Z}(x,\ell;\theta)}{\eta_{Z}^{( I, \theta )}(\ell)} \, , \\
	\psi_{Z}( x, \ell; \theta) & =	\begin{cases}
							\dfrac{P_{D}(x,\ell) \, g(z_{\theta(\ell)}|x,\ell)}{\kappa(z_{\theta(\ell)})}, & \mbox{if}\ \theta(\ell)>0\\
							1-P_{D}(x,\ell), & \mbox{if}\ \theta(\ell)=0
						\end{cases} \, . 
\end{align}
Using now (\ref{eq:mdglmb_w})-(\ref{eq:mdglmb_p}), the M$\delta$-GLMB density corresponding to the $\delta$-GLMB density in (\ref{eq:mdglmbupdatedpdf}) is a probability density of the form (\ref{eq:mdglmb}) with
\begin{align}
	w^{(I)} & = \sum_{\theta\in\Theta(I)}w^{(I,\theta)}\label{eq:mdglmbweight}\\
	p^{(I)}(x,\ell) & = 1_{I}(\ell)\frac{1}{w^{(I)}}\sum_{\theta\in\Theta(I)}w^{(I,\theta)}p^{(I,\theta)}(x,\ell)\label{eq:mdglmbpdf}
\end{align}
The M$\delta$-GLMB density provided by (\ref{eq:mdglmbweight})-(\ref{eq:mdglmbpdf}) preserves both PHD and cardinality distribution of the original $\delta$-GLMB density.
Summing up, at each time instant $k$, (\ref{eq:mdglmbupdatedpdf})-(\ref{eq:mdglmbpdf}) have to be carried sequentially for each sensor $s$ to evaluate (\ref{eq:msMTBayesUpdate}).

\begin{rem}\label{rem:maxhpup}
	Each hypothesis $I \in \mathcal{F}\left( \mathbb{L} \right)$ generates a set of $\left| \Theta(I) \right|$ new measurement-to-track association maps for the $\delta$-GLMB posterior.
	The number of components $\left( w_{k+1|k}^{\left(I, \theta\right)}, p_{k+1|k}^{\left(I, \theta\right)} \right)$ stored/computed after the M$\delta$-GLMB update step (\ref{eq:mdglmbupdatedpdf}) is $\left| \mathcal{F}\!\left( \mathbb{L} \right) \right| \cdot \sum_{I \in \mathcal{F}\left(\mathbb{L}\right)}\left| \Theta(I) \right|$.
	On the other hand, the number of hypotheses $\left( w_{k+1|k}^{\left(I, \xi, \theta \right)}, p_{k+1|k}^{\left(\xi, \theta\right)} \right)$ after the $\delta$-GLMB update (\ref{eq:dglmbupdatedpdf}) is $\left| \mathcal{F}\!\left( \mathbb{L} \right) \times \Xi \right| \cdot\sum_{I \in \mathcal{F}(\mathbb{L})}\left| \Theta(I) \right|$ for $w_{k+1|k}^{\left(I, \xi, \theta \right)}$ and $\left| \Xi \right| \cdot \sum_{I \in \mathcal{F}(\mathbb{L})}\left| \Theta(I) \right|$ for $p_{k+1|k}^{\left(\xi, \theta\right)}$.
	The same conclusions along the lines of Remark \ref{rem:maxhppred} hold.
\end{rem}

\begin{rem}\label{rem:maxhpmarginal}
	After the marginalization procedure (\ref{eq:mdglmbweight})-(\ref{eq:mdglmbpdf}) only $\left|\mathcal{F}(\mathbb{L})\right|$ hypotheses are retained, as all the new contributions provided by the association maps $\left| \Theta(I) \right|$ are aggregated in a single component.
	Notice that $\left|\mathcal{F}(\mathbb{L})\right|$ is the exact same number of hypotheses produced during the prediction step (\ref{eq:mdglmbpredictedpdf}) (see Remark \ref{rem:maxhppred}).
	Thus, the prediction step (\ref{eq:mdglmbupdatedpdf}) sets the upper bound of the total hypotheses that will be retained after each full M$\delta$-GLMB step.
\end{rem}

From Remark \ref{rem:maxhpup} and \ref{rem:maxhpmarginal}, the M$\delta$-GLMB is preferable over the $\delta$-GLMB in terms of stored information and computational burden, since the number of remaining hypotheses after each sensor update step in (\ref{eq:msMTBayesUpdate}) is always set to $\left|\mathcal{F}(\mathbb{L})\right|$.
Note that this does not apply to the $\delta$-GLMB due to the super-exponential growth as reported in Remark \ref{rem:maxhppred}.
This is an important property of the M$\delta$-GLMB since it yields a principled approximation which greatly decreases the need of pruning hypotheses w.r.t. the $\delta$-GLMB \cite{vovo2}.
In fact, pruning in the $\delta$-GLMB might lead to poor performance in multi-sensor scenarios with low SNR (e.g. high clutter intensity, low probability of detection, etc.) and limited storage/computational capabilities.
For instance, this may happen if a subset of the sensors do not detect one or more targets and hypotheses associated to the true tracks are removed due to pruning.
Furthermore, from a mathematical viewpoint, pruning between corrections generally produces a less informative and order-independent approximation to the posterior distribution in eq. (\ref{eq:msMTBayesUpdate}).

\subsection{M$\delta$-GLMB Filter Implementation}
The $\delta$-GLMB filter implementation \cite{vovo2} applies directly to M$\delta$-GLMB.
For a linear Gaussian multi-target model it is assumed that \textsc{i}) the single target transition density, likelihood and birth intensity are assumed to be Gaussian; \textsc{ii}) survival and detection probabilities are constants; \textsc{iii}) each single target density is represented as a Gaussian mixture.
The corresponding Gaussian mixture predicted and updated densities are computed using the standard Gaussian mixture update and prediction formulas based on the Kalman filter \cite{vovo2}.
In the case of having non-linear single target transition density and/or likelihood, one can resort to the well known Extended or Unscented Kalman Filters \cite{ekf,ukf}.
On the other hand, for non-linear non-Gaussian multi-target models (with state dependent survival and detection probabilities), each single target density can be represented by a set of weighted particles.
The corresponding predicted and updated densities are computed by the standard particle (or Sequential Monte Carlo) filter \cite{smc,VSD05,ristic}.

\subsection{Connection with the LMB Filter}
The LMB filter introduced in \cite{lmbf} is a single component approximation
to a $\delta$-GLMB density that matches the unlabeled PHD. In this
subsection we show an alternative derivation of the LMB approximation
first proposed in \cite{lmbf} through a connection with the M$\delta$-GLMB
approximation. Recall that a LMB density is uniquely parameterized
by a set of existence probabilities $r^{\left(\ell\right)}$ and corresponding
track densities $p^{(\ell)}(\cdot)$:
\begin{equation}
	\boldsymbol{\pi}_{LMB}(\mathbf{X})=\Delta(\mathbf{X})w(\mathcal{L}(\mathbf{X}))p^{\mathbf{X}}\label{eq:lmbpdf}
\end{equation}
where
\begin{align}
	w(L) & =\prod_{\imath\in\mathbb{L}}\left(1-r^{\left(\imath\right)}\right)\prod_{\ell\in L}\frac{1_{\mathbb{L}}(\ell)r^{\left(\ell\right)}}{1-r^{(\ell)}}\\
	p(x,\ell) & =p^{(\ell)}(x)
\end{align}

In the following we show that by extracting individual tracks from
the M$\delta$-GLMB approximation, we can obtain the same expressions
for the existence probabilities and state densities originally proposed
for the LMB filter in \cite{lmbf}:
\begin{align}
	r^{(\ell)} & =\mbox{Pr}_{\hat{\boldsymbol{\pi}}}(\mathcal{L}(\mathbf{X})\ni\ell)=\sum_{L\ni\ell}w(L)\\
	& = \sum_{I\in\mathcal{F}(\mathbb{L})}1_{I}(\ell)w(I)\\
	& = \sum_{I \in \mathcal{F}(\mathbb{L})}1_{I}(\ell)w^{\left(I\right)}\\
	& = \sum_{(I, \xi) \in \mathcal{F}(\mathbb{L}) \times \Xi}1_{I}(\ell)w^{\left(I, \xi\right)}
\end{align}
\begin{align}
	p^{(\ell)}(x) & =\dfrac{\mbox{Pr}_{\hat{\boldsymbol{\pi}}}(\mathbf{X}\ni(x,\ell))}{\mbox{Pr}_{\hat{\boldsymbol{\pi}}}(\mathcal{L}(\mathbf{X})\ni\ell)}\label{eq:defphd} \\
	& =\dfrac{\displaystyle\int\hat{\boldsymbol{\pi}}((x,\ell)\cup\mathbf{X})\delta\mathbf{X}}{r^{(\ell)}}\label{eq:phd}
\end{align}
where the notation for the numerator in (\ref{eq:defphd}) is defined as per \cite[eq. 11.111]{Mahler07}, while the numerator of (\ref{eq:phd}) follows from \cite[eq. 11.112]{Mahler07}.
Notice that the numerator is precisely the PHD $\hat{v}$ corresponding to $\hat{\boldsymbol{\pi}}$, which by Proposition 2 of \cite{GLMBapprox} exactly matches the PHD $v$ corresponding to $\pi$.
Using the results in \cite{GLMBapprox}, it can be verified that 
\begin{equation}
	\hat{v}(x,\ell) = \sum_{I \in \mathcal{F}(\mathbb{L})} 1_{I}(\ell) w^{(I)}p^{(I)}(x,\ell)
\end{equation}
and consequently
\begin{align}
	p^{(\ell)}(x) & = \frac{\hat{v}(x,\ell)}{r^{(\ell)}} = \frac{1}{r^{(\ell)}}\sum_{I \in \mathcal{F}(\mathbb{L})} 1_{I}(\ell) w^{(I)}p^{(I)}(x,\ell)\\
	& = \frac{1}{r^{(\ell)}}\sum_{(I, \xi) \in \mathcal{F}(\mathbb{L}) \times \Xi} 1_{I}(\ell) w^{(I, \xi)}p^{(\xi)}(x,\ell)
\end{align}

Notice that, however, the property of matching the labeled PHD of the $\delta$-GLMB does not hold for the LMB filter, as shown in \cite[Section III]{lmbf}, due to the imposed multi-Bernoulli structure for the cardinality distribution.

\section{Numerical Results}
\label{sec:results}
\begin{figure}[h!]
        \begin{minipage}[t][][t]{0.475\columnwidth}
		\includegraphics[width=\columnwidth]{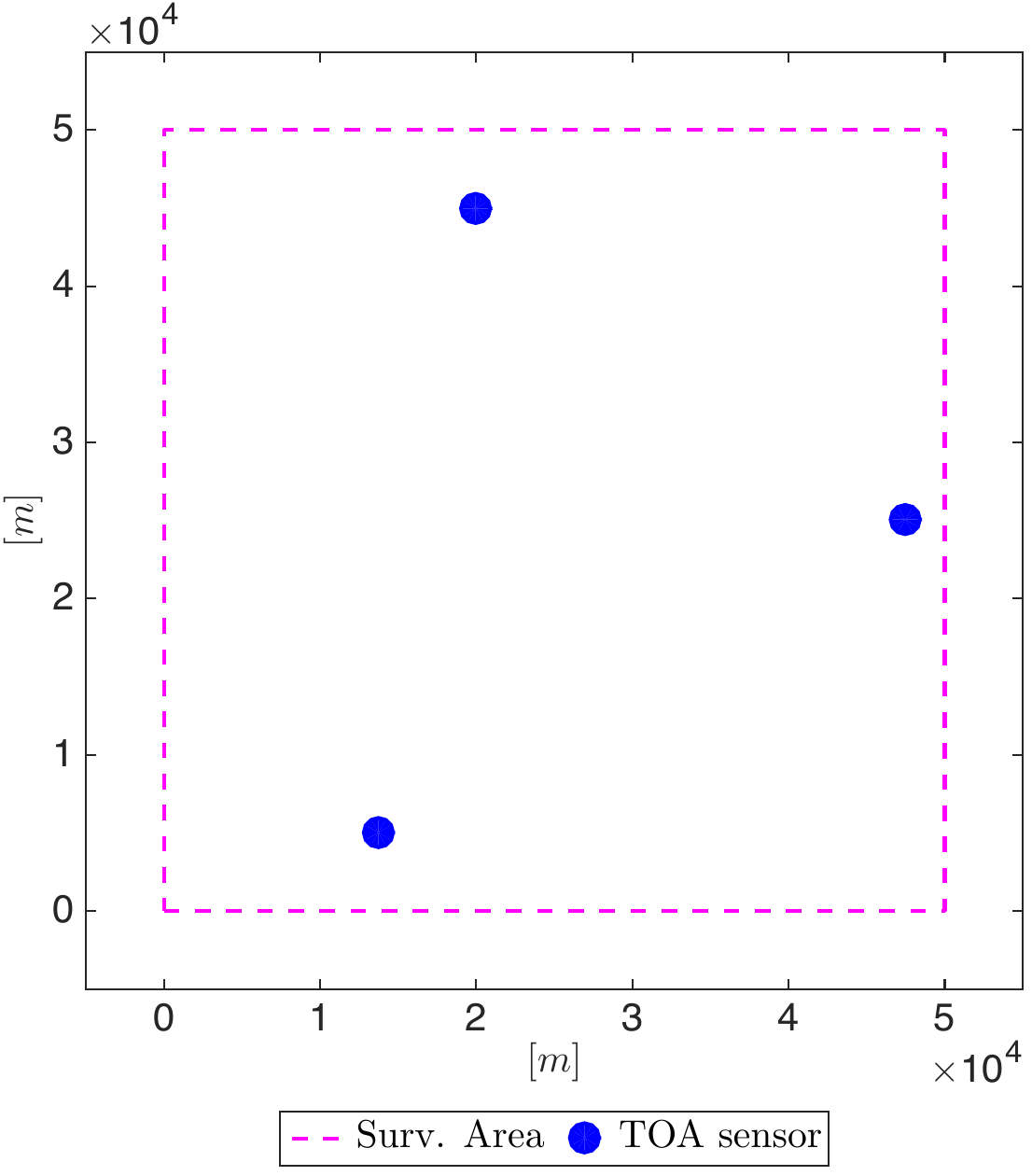}
		\caption{Network with 3 TOA sensors.}
		\label{fig:3toa}
        \end{minipage}
        \begin{minipage}[t][][t]{0.475\columnwidth}
		\includegraphics[width=\columnwidth]{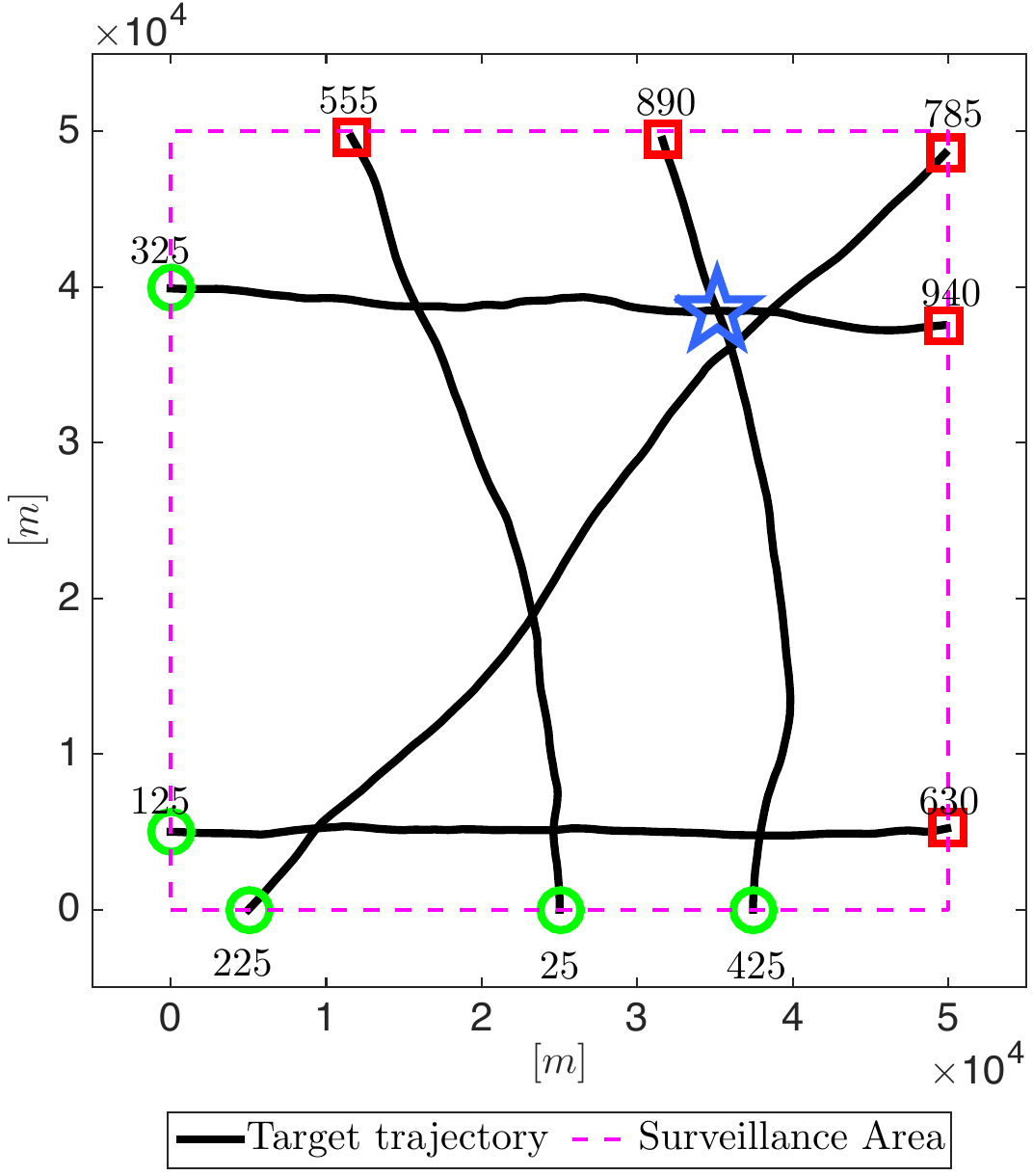}
		\caption{Target trajectories considered in the simulation experiment. The start/end point for each trajectory is denoted, respectively, by $\bullet\backslash\blacksquare$. The {\Large$\star$} indicates a rendezvous point.}
		\label{fig:5trajectories}
        \end{minipage}
\end{figure}

To assess performance of the proposed Marginalized $\delta$-GLMB (M$\delta$-GLMB), a $2$-dimensional multi-object tracking scenario is considered over a surveillance area of $50\times50 \, [km^{2}]$.
Two sensor sets are used to represent scenarios with different observability capabilities.
In particular:
\textsc{i}) a single Radar in the middle of the surveillance region is used as it guarantee observability;
\textsc{ii}) a set of $3$ \textit{range-only} (Time Of Arrival, TOA), deployed as shown in Fig. \ref{fig:3toa}, are used as they do not guarantee observability individually, but information from different sensors need to be combined to achieve it.

The scenario consists of $5$ targets as depicted in Fig. \ref{fig:5trajectories}.
For the sake of comparison, the M$\delta$-GLMB is also compared with the $\delta$-GLMB ($\delta$-GLMB) \cite{vovo1,vovo2} and LMB (LMB) \cite{lmbf} filters.
The three tracking filters are implemented using Gaussian Mixtures to represent their predicted and updated densities \cite{vovo2,lmbf}.
Due to the non linearity of the sensors, the \textit{Unscented Kalman Filter} (UKF) \cite{juluhl2004} is exploited to update means and covariances of the Gaussian components.

The kinematic object state is denoted by $x = \left[ p_{x}, \, \dot{p}_{x}, \, p_{y}, \, \dot{p}_{y} \right]^{\top}$, i.e. the planar position and velocity.
The motion of objects is modeled according to the Nearly-Constant Velocity (NCV) model \cite{farina1,farina2,barshalom1,barshalom2}:
\begin{equation*}
x_{t + 1} = \left[ \begin{array}{cccc}
1 & T_{s} & 0 & 0	\\
0 & 1 	  & 0 & 0		\\
0 & 0 	  & 1 & T_{s} \\
0 & 0 	  & 0 & 1		\end{array} \right] x_{t} + w_{t} \, , \qquad
Q = \sigma_{w}^{2} \left[ \begin{array}{cccc}
\frac{1}{4}T_{s}^{4} & \frac{1}{2}T_{s}^{3} & 0 & 0 \\
\frac{1}{2}T_{s}^{3} & T_{s}^{2} & 0 & 0 \\
0 & 0 & \frac{1}{4}T_{s}^{4} & \frac{1}{2}T_{s}^{3}\\
0 & 0 & \frac{1}{2}T_{s}^{3} & T_{s}^{2} \end{array} \right]
\end{equation*}
where $\sigma_{w} = 5 \, [m/s^{2}]$ and the sampling interval is $T_{s} = 5\,[s]$.

The Radar has the following measurement function:
\begin{equation}
\begin{array}{c}
h(x) = \left[ \begin{array}{ll}
				\angle [ \left( p_{x} - x^{r} \right) + j \left( p_{y} - y^{r} \right)] \\[0.5em]
                                	\sqrt{ \left( p_{x} - x^{r} \right)^2+ \left( p_{y} - y^{r} \right)^2 }
			\end{array} \right]
\end{array}
\end{equation}
where $( x^{r}, y^{r} )$ represents the known position of the Radar and its measurement noise is $\sigma_{Radar} = \left[ 1 \, [\mbox{}^{\circ}] \, , 100 \, [m]\right]^{\top}$.
The measurement functions of the $3$ TOA of Fig. \ref{fig:3toa} are:
\begin{equation}
h(x) = \sqrt{ \left( p_{x} - x^{s} \right)^2+ \left( p_{y} - y^{s} \right)^2 } \, ,
\end{equation}
where $( x^{s}, y^{s} )$ represents the known position of sensor (indexed with) $s$. The standard deviation of the TOA measurement noise is taken as $\sigma_{TOA} = 100 \, [m]$.

The clutter is characterized by a Poisson process with parameter $\lambda_{c} = 15$.
The probability of target detection is $P_{D} = 0.85$.

In the considered scenario, targets pass through the surveillance area with partial prior information for target birth locations. 
Accordingly, a $10$-component LMB RFS $\boldsymbol{\pi}_{B} = \left\{ \left( r^{\left( \ell \right)}_{B}, p^{\left( \ell \right)}_{B} \right) \right\}_{\ell \in \mathbb{B}}$ has been hypothesized for the birth process.
Table \ref{tab:borderlineinit} gives detailed summary of such components.
\begin{table}[h!]
	\setlength\arrayrulewidth{0.5pt}\arrayrulecolor{black} 
	\setlength\doublerulesep{0.5pt}\doublerulesepcolor{black} 
	\caption{Components of the LMB RFS birth process at a given time $k$.}
	\label{tab:borderlineinit}
	\centering
	$r^{\left( \ell \right)} = 0.09$\\
	$p^{\left( \ell \right)}_{B}(x) = \mathcal{N}\!\left( x;\, m^{\left( \ell \right)}_{B}, P_{B} \right)$\\
	$P_{B} = \operatorname{diag}\!\left( 10^{6}, 10^{4}, 10^{6}, 10^{4} \right)$\\\vspace{0.5em}
	\scalebox{1}{
	\begin{tabular}{>{\columncolor[gray]{.95}}c||c|c|c||}
		{\textbf{Label}} & $\left( k, \, 1 \right)$ & $\left( k, \, 2 \right)$ & $\left( k, \, 3 \right)$ \\
		\hline
		$m^{\left( \ell \right)}_{B}$ & $\left[ 0, \, 0, \, 40000,\, 0 \right]^{\top}$ & $\left[ 0, \, 0, \, 25000,\, 0 \right]^{\top}$ & $\left[ 0, \, 0, \, 5000,\, 0 \right]^{\top}$\\
		\hline
		\hline
	\end{tabular}
	}\\\vspace{0.5em}
	\scalebox{1}{
	\begin{tabular}{>{\columncolor[gray]{.95}}c||c|c|c||}
		{\textbf{Label}} & $\left( k, \, 4 \right)$ & $\left( k, \, 5 \right)$ & $\left( k, \, 6 \right)$\\
		\hline
		$m^{\left( \ell \right)}_{B}$ & $\left[ 5000, \, 0, \, 0,\, 0 \right]^{\top}$ & $\left[ 25000, \, 0, \, 0,\, 0 \right]^{\top}$ & $\left[ 36000, \, 0, \, 0,\, 0 \right]^{\top}$\\
		\hline
		\hline
	\end{tabular}
	}\\\vspace{0.5em}
	\scalebox{1}{
	\begin{tabular}{>{\columncolor[gray]{.95}}c||c|c||}
		{\textbf{Label}} & $\left( k, \, 7 \right)$ & $\left( k, \, 8 \right)$\\
		\hline
		$m^{\left( \ell \right)}_{B}$ & $\left[ 50000, \, 0, \, 15000,\, 0 \right]^{\top}$ & $\left[ 50000, \, 0, \, 40000,\, 0 \right]^{\top}$\\
		\hline
		\hline
	\end{tabular}
	}\\\vspace{0.5em}
	\scalebox{1}{
	\begin{tabular}{>{\columncolor[gray]{.95}}c||c|c||}
		{\textbf{Label}} & $\left( k, \, 9 \right)$ & $\left( k, \, 10 \right)$\\
		\hline
		$m^{\left( \ell \right)}_{B}$ & $\left[ 40000, \, 0, \, 50000,\, 0 \right]^{\top}$ & $\left[ 10000, \, 0, \, 50000,\, 0 \right]^{\top}$\\
		\hline
		\hline
	\end{tabular}
	}
\end{table}
Due to the partial prior information on the object birth locations, some of the LMB components cover a state space region where there is no birth. Therefore, clutter measurements are more prone to generate false targets.

Multi-target tracking performance is evaluated in terms of the \textit{Optimal SubPattern Analysis} (OSPA) metric \cite{schvovo2008} with Euclidean distance, $p = 2$, and cutoff $c = 600 \, [m]$.
The reported metric is averaged over $100$ Monte Carlo trials for the same target trajectories but different, independently generated, clutter and measurement noise realizations. The duration of each simulation trial is fixed to $1000 \, [s]$ ($200$ samples).

The three tracking filters are coupled with the \textit{parallel CPHD look ahead strategy} described in \cite{vovo1,vovo2}. The CPHD \cite{vo-vo-cantoni} filter.

\subsection{Scenario 1: Radar}
\begin{figure}[h!]
        \begin{minipage}[t][][t]{\columnwidth}
	        	\centering
		\includegraphics[width=\columnwidth]{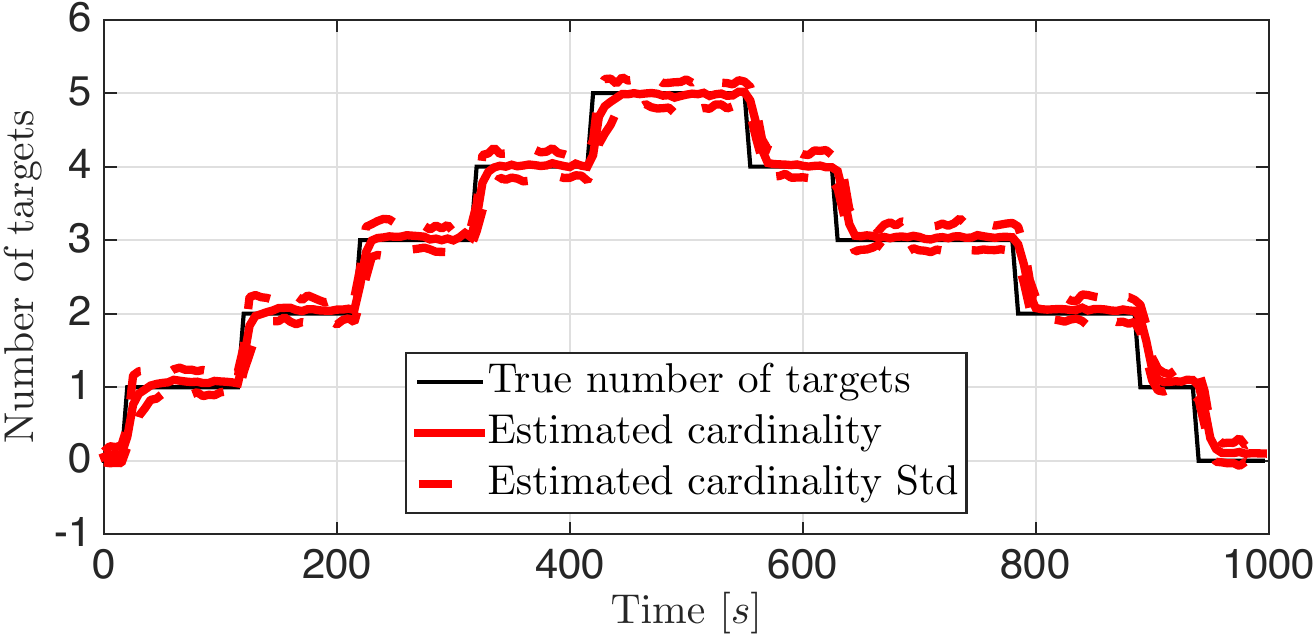}
		\caption{Cardinality statistics for M$\delta$-GLMB tracking filter using 1 Radar.}
		\label{fig:1:cardMDGLMB}
        \end{minipage}\vspace{0.5em}
        \begin{minipage}[t][][t]{\columnwidth}
	        \centering
		\includegraphics[width=\columnwidth]{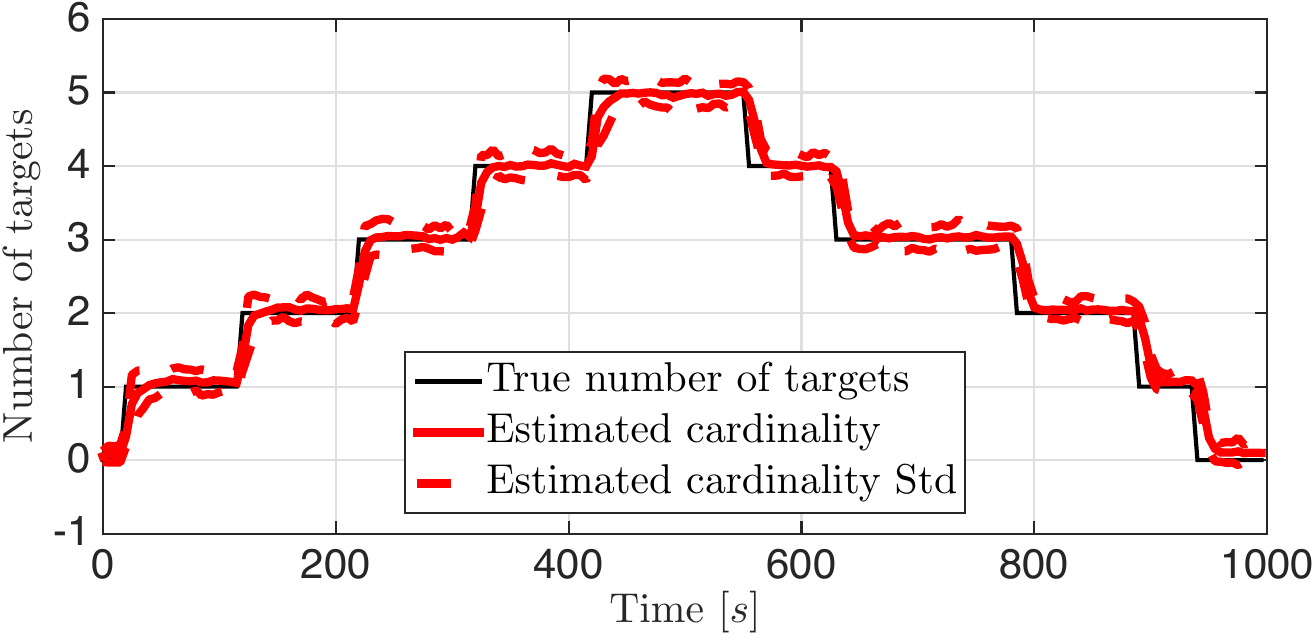}
		\caption{Cardinality statistics for $\delta$-GLMB tracking filter using 1 Radar.}
		\label{fig:1:cardDGLMB}
        \end{minipage}
\end{figure}

\begin{figure}[h!]
        \begin{minipage}[t][][t]{\columnwidth}
	        	\centering
		\includegraphics[width=\columnwidth]{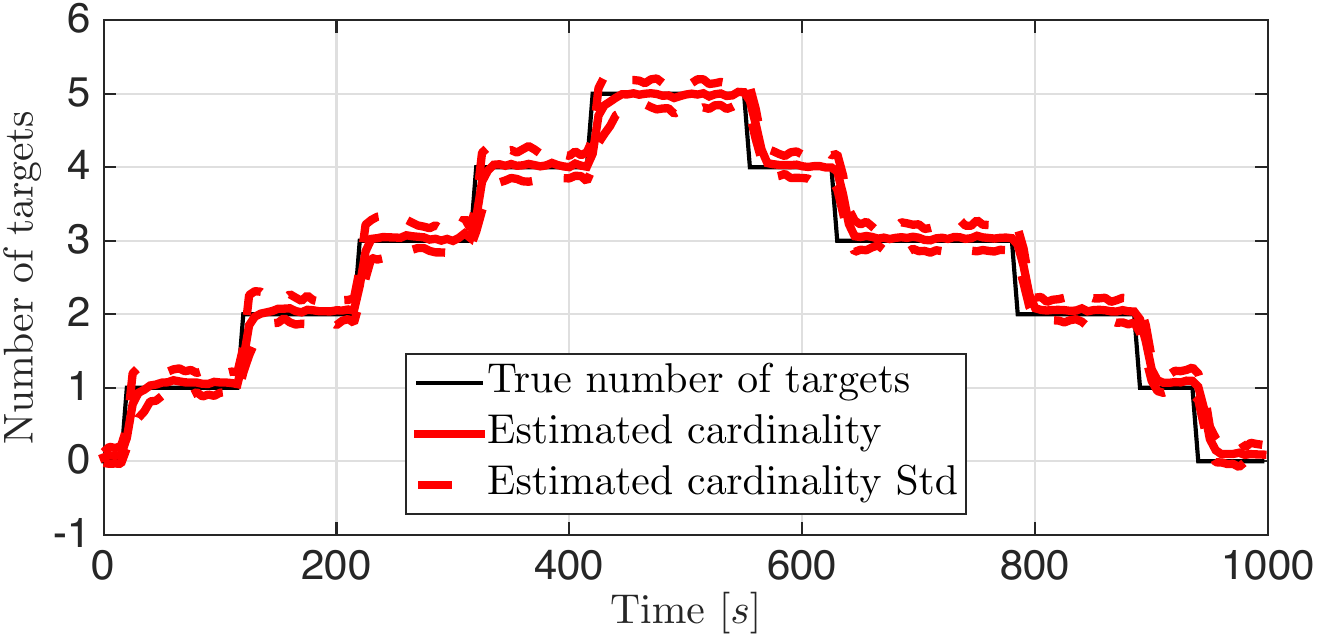}
		\caption{Cardinality statistics for LMB tracking filter using 1 Radar.}
		\label{fig:1:cardLMB}
        \end{minipage}\vspace{0.5em}
        \begin{minipage}[t][][t]{\columnwidth}
        		\centering
		\includegraphics[width=\columnwidth]{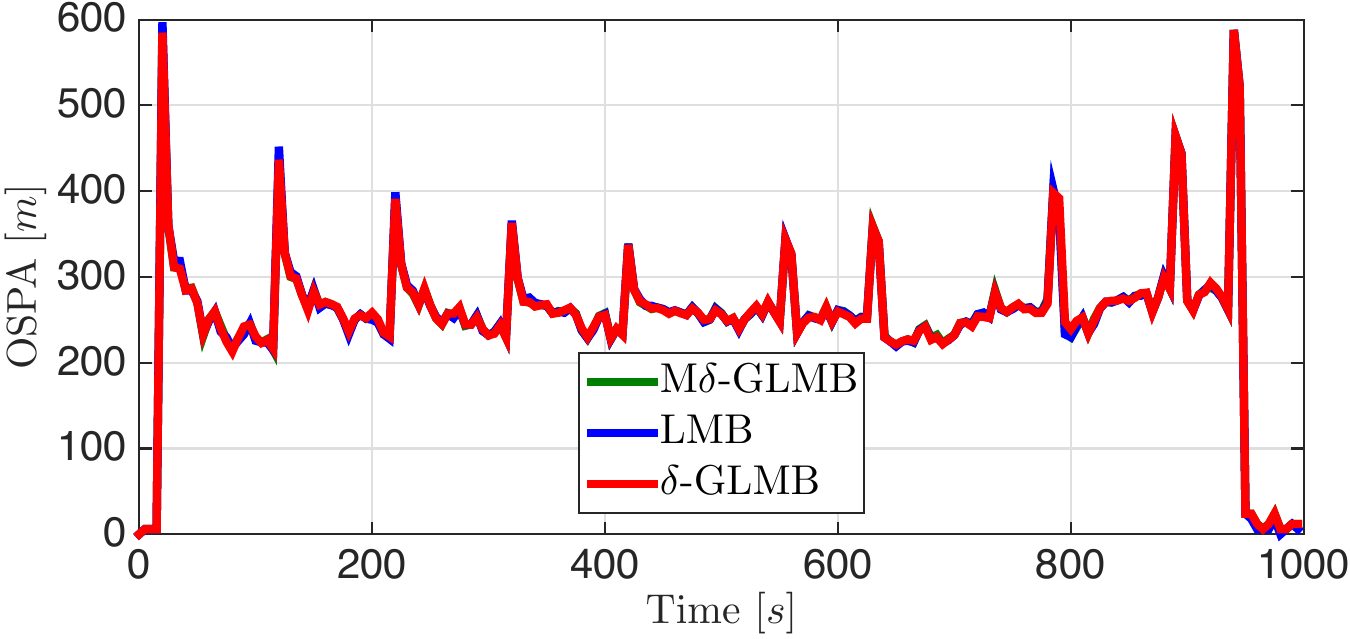}
		\caption{OSPA distance ($c = 600 \, [m]$, $p = 2$) using 1 Radar.}
		\label{fig:1:ospa}
        \end{minipage}
\end{figure}

Figs. \ref{fig:1:cardMDGLMB}, \ref{fig:1:cardDGLMB} and \ref{fig:1:cardLMB} display the statistics (mean and standard deviation) of the estimated number of targets obtained, respectively, with the M$\delta$-GLMB, the $\delta$-GLMB and the LMB.
As it can be seen, all the algorithms estimate the target cardinality accurately, with no substantial differences.
This result indicates that, in the presence of a single sensor guaranteeing observability, the approximations made by both M$\delta$-GLMB and LMB are not critical in that they provide performance comparable to the $\delta$-GLMB with the advantage of a cheaper computational burden and reduced storage requirements.
Note that the problems introduced by the rendezvous point (e.g. merged or lost tracks) are correctly tackled by all the algorithms.

Fig. \ref{fig:1:ospa} shows the OSPA distance of the algorithms.
Note again that, in agreement with the estimated cardinality distributions, the OSPA distances are nearly identical.

\subsection{Scenario 2: 3 TOA}
\begin{figure}[h!]
	\begin{minipage}[t][][t]{\columnwidth}
		\centering
		\includegraphics[width=\columnwidth]{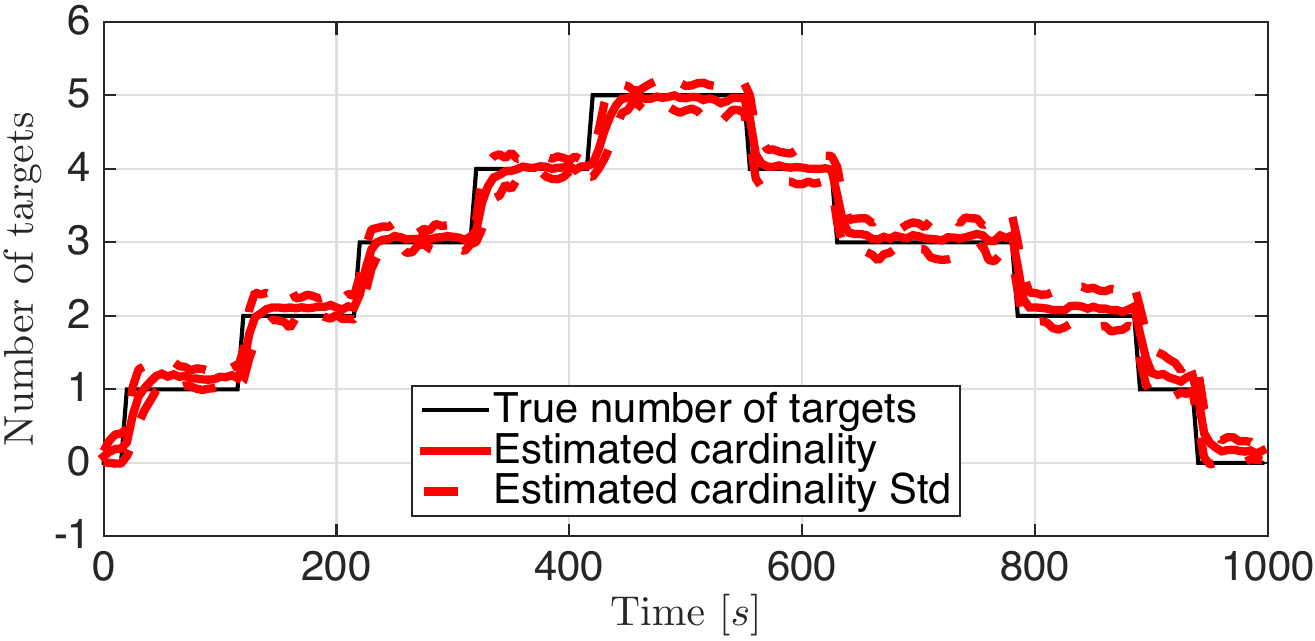}
		\caption{Cardinality statistics for M$\delta$-GLMB tracking filter using 3 TOA.}
		\label{fig:2:cardMDGLMB}
        \end{minipage}\vspace{0.5em}
        \begin{minipage}[t][][t]{\columnwidth}
        		\centering
		\includegraphics[width=\columnwidth]{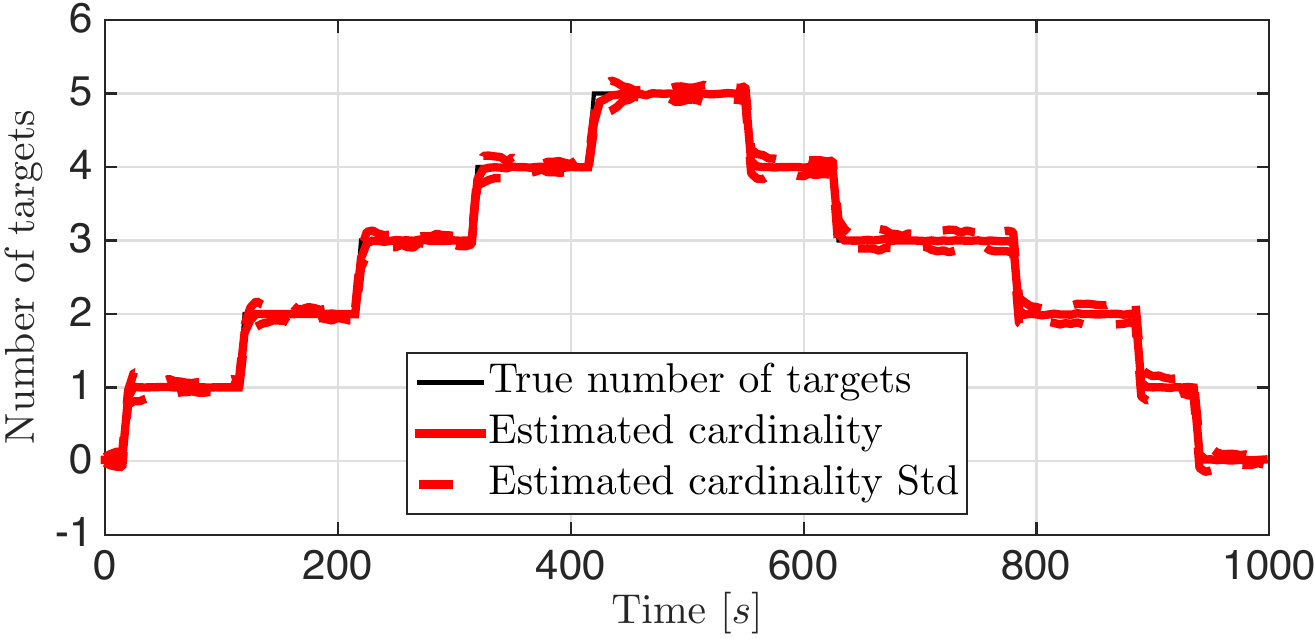}
		\caption{Cardinality statistics for $\delta$-GLMB tracking filter using 3 TOA.}
		\label{fig:2:cardDGLMB}
        \end{minipage}
\end{figure}

\begin{figure}[h!]
        \begin{minipage}[t][][t]{\columnwidth}
        		\centering
		\includegraphics[width=\columnwidth]{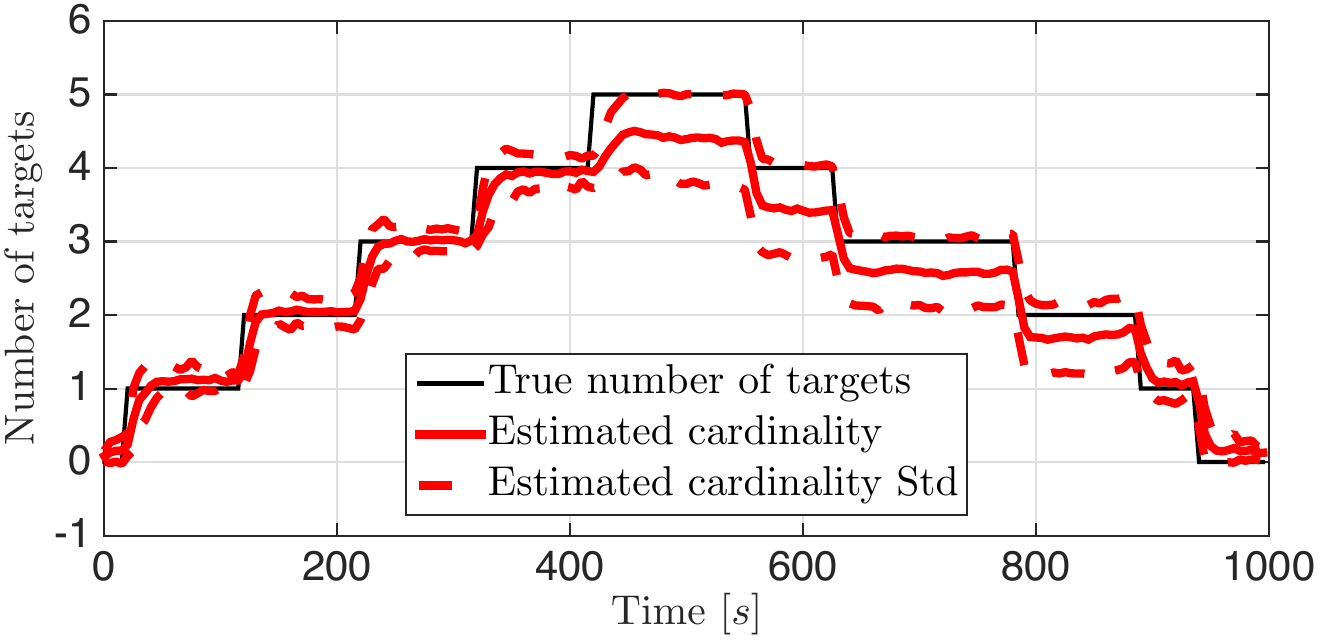}
		\caption{Cardinality statistics for LMB tracking filter using 3 TOA.}
		\label{fig:2:cardLMB}
        \end{minipage}\vspace{0.5em}
        \begin{minipage}[t][][t]{\columnwidth}
        		\centering
		\includegraphics[width=\columnwidth]{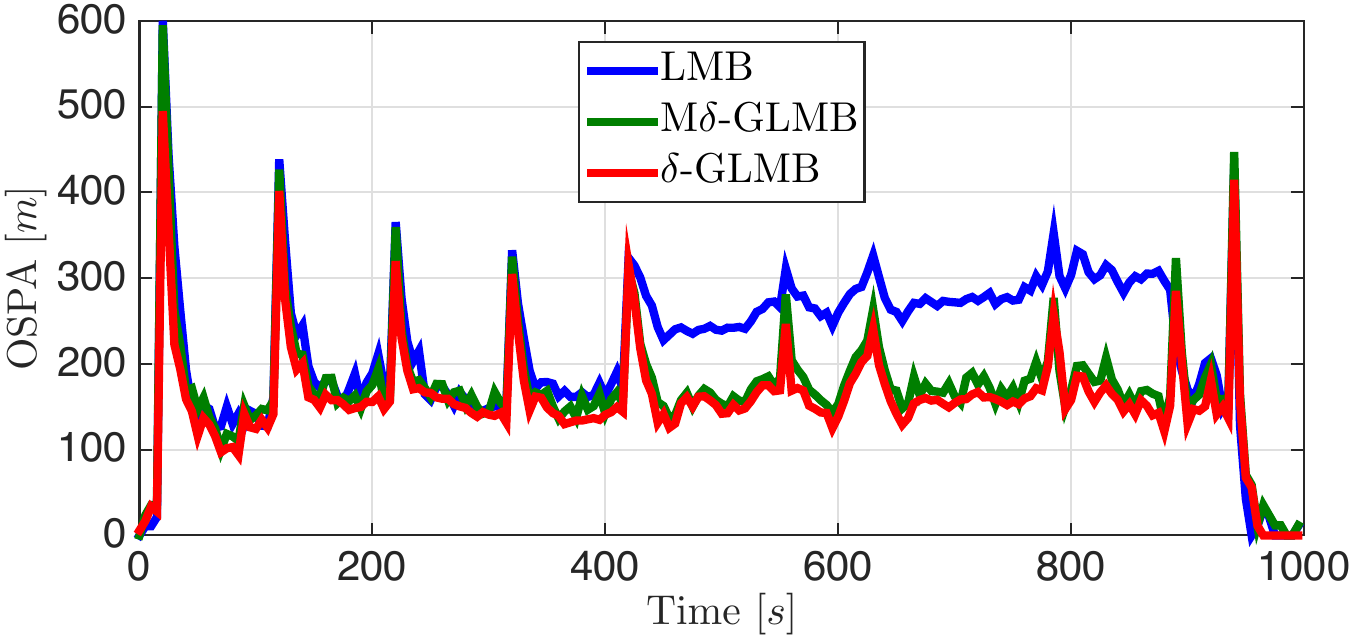}
		\caption{OSPA distance ($c = 600 \, [m]$, $p = 2$) using 3 TOA.}
		\label{fig:2:ospa}
        \end{minipage}
\end{figure}

Figs. \ref{fig:2:cardMDGLMB}, \ref{fig:2:cardDGLMB} and \ref{fig:2:cardLMB} display the statistics (mean and standard deviation) of the estimated number of targets obtained, respectively, with the M$\delta$-GLMB, the $\delta$-GLMB and the LMB.
The M$\delta$-GLMB and the $\delta$-GLMB tracking filters estimate the target cardinality accurately, while the LMB exhibits poor performance and higher standard deviation due to losing some tracks when 4 or 5 targets are jointly present in the surveillance area.
It is worth noticing that the M$\delta$-GLMB performs as nearly as identical to the $\delta$-GLMB and that the problems introduced by the rendezvous point are again correctly tackled.

Fig. \ref{fig:2:ospa} shows the OSPA distance.
Note that the OSPA of the M$\delta$-GLMB is close to the one of $\delta$-GLMB, while the LMB shows an overall higher error in agreement with the cardinality error due to losing tracks.

\clearpage{}

\section{Conclusion and future work}
This paper has proposed a novel approximation to the $\delta$-GLMB filter with standard point detection measurements.
The result is based on a principled GLMB approximation to the labeled RFS posterior that matches exactly the posterior PHD and cardinality distribution.
The proposed approximation can be interpreted as performing a marginalization with respect to the association histories arising from the $\delta$-GLMB filter.
The key advantage of the new filter lies in the reduced growth rate of the number of new components generated at each filtering step.
In particular, the approximation (or marginalization) step performed after each update is guaranteed to reduce the number of generated components which normally arise from multiple measurement-to-track association maps.
Typically, the proposed M$\delta$-GLMB filter requires much less computation and storage especially in multi-sensor scenarios compared to the $\delta$-GLMB filter.
Furthermore the proposed M$\delta$-GLMB filter inherits the same implementation strategies and parallelizability of the $\delta$-GLMB filter.
A connection and alternative derivation of the LMB filter is also provided.
Future works will consider distributed estimation with the M$\delta$-GLMB filter.

\clearpage{}

\end{document}